# WDX-Analysis of the New Superconductors $R$O$_{1-x}$F$_x$FeAs and Its Consequences on the Electronic Phase Diagram

Anke Köhler, Günter Behr

*IFW Dresden, P.B. 270116 Dresden, Germany*
e-mail: anke_koehler@yahoo.de

Polycrystalline samples of $R$O$_{1-x}$F$_x$FeAs ($0 \leq x \leq 0.25$) ($R$ = La, Sm, Gd) were investigated by wavelength-dispersive X-ray spectroscopy (WDX) in the electron microscope to determine the composition of the samples, in particular the fluorine content. It was found that the measured fluorine content can deviate considerably from the initial weight. In the lanthanum compound LaO$_{1-x}$F$_x$FeAs, we found good agreement mainly for $x \geq 0.05$, but for $x < 0.05$ the fluorine hardly goes into the sample. For the samarium compound we measured less than half the fluorine in the sample as initially weighed at all fluorine concentrations. These measured values are taken into account when drawing the electronic phase diagrams of LaO$_{1-x}$F$_x$FeAs and SmO$_{1-x}$F$_x$FeAs. This leads to a more consistent picture of both of the diagrams in comparison to the plot of the initial weight.

## 1 Introduction

Although much progress has been made in the study of $R$O$_{1-x}$F$_x$FeAs compounds, there are still contradictory opinions about their physical properties: for instance, the fluorine doping level at which superconductivity occurs and the doping level at which the maximum transition temperature T$_c$ is obtained. Also there is disagreement as to whether a first- or second-order transition takes place between the magnetic and superconducting region of the electronic phase diagram. When comparing data from the literature (see e.g. Uemura [1]) it is obvious that the electronic phase diagrams of Sm, Ce and La look different. And it is not only in the case of different rare earth elements that the data differ, but also for La or Sm itself one can find plenty of varying data.

In most of the data reported in the literature the fluorine content was not measured at all, but only the initial weight was used for plotting the diagrams. For us, over many years of sample preparation it consistently proved true that it is necessary to analyze the composition after preparing the sample.

It should also be mentioned that it is not only the fluorine content that determines the physical properties of these compounds, but also parameters such as oxygen [2] or arsenic deficiencies [3], structural changes, and inhomogeneities [4]. However in this article the focus will be on the measurement of the fluorine content.

The method we used is electron beam microprobe analysis, which is local (μm-sized) and non-destructive. The electron microscope is equipped with both an energy-dispersive X-ray (EDX) detector and a wavelength-dispersive X-ray (WDX) detector. The advantages of the WDX detector are its better spectral resolution and its ability to analyze light elements down to the element boron routinely. But one has to accept a long measurement time. While the EDX measurements can be performed without application of any standard materials, the WDX analysis needs the use of standards. In our case, we used $Fe_2O_3$ and the rare earth fluorides $RF_3$ ($R$ = La, Sm, Gd), as well as single-crystalline FeAs. We found the largest measurement uncertainty comes from the sample itself, since the fluorine content varies locally inside the polycrystalline sample. Thus we decided to measure approximately ten different points and take the root mean square deviation as the measurement error. To minimize the unknown absorption error of the low energy radiation (of light elements), we normalized the fluorine concentration $c_F$ on the oxygen concentration $c_O$ and used the ratio $c_F / (c_O + c_F)$ for the comparison with the initial value x.

## 2 Results

The samples consist mainly of the $RO_{1-x}F_xFeAs$ phase; only small amounts of secondary phases like $RO_yF_z$ and FeAs were found. The determined fluorine concentrations in the main phase of $SmO_{1-x}F_xFeAs$ and $LaO_{1-x}F_xFeAs$ are plotted in Figs. 1 and 2, respectively.

For the samarium compound, we found a large deviation of the measured fluorine concentrations from the initial weight. Only less than the half of the initially weighed fluorine was incorporated in the main phase. The lack of fluorine in the main phase can be caused by fluorine incorporation in the fluorine rich secondary $RO_yF_z$ phase, by the formation of volatile components, or by absorption from the enveloping material. Although the reason is still unclear, we found some conditions which support or prevent fluorine incorporation. Using different grinding procedures for the Gd compound with the nominal composition $GdO_{0.83}F_{0.17}FeAs$, it turned out that a fine-grained powder has better potential for fluorine incorporation. On the other hand, fluorine absorption is suppressed in the required phase by any oversupply of oxygen.

In the lanthanum compound, the measured fluorine values are in better agreement with the initial weight. But a substantial deviation is seen at $x < 0.05$. The point at $x = 0.03$, which corresponds to the initial value, originates from an As-deficient sample and therefore is less representative. Another outward lying point is seen at $x = 0.15$.

Figures 3 and 4 show the electronic phase diagrams of $SmO_{1-x}F_xFeAs$ and $LaO_{1-x}F_xFeAs$, respectively. The critical temperature data originate from magnetic and resistivity measurements ($\chi(T)$, $\mu SR$, $R(T)$; see [5],[6]). In these figures, the data are plotted against both the nominal fluorine content (small symbols, thin line) and the measured fluorine content (larger symbols, bold line). Moving from nominal to measured values of the fluorine content all data points for the Sm compound shift to the left (Figure 3). A similar shift occurs for $x < 0.05$ in the La compound (Figure 4). For $x \geq 0.05$, we stop plotting the new points due to the good agreement between nominal and measured fluorine values (except for $x = 0.15$). In the region of 4% fluorine content, we cannot yet provide a data point. Thus, the precise appearance of the transition region between the magnetic and the superconducting part of the phase diagram is not completely clear. But we expect there is no abrupt transition from magnetic to superconducting order. When comparing the two phase diagrams (Figures 3 and 4), one sees good agreement in the region between 0 % and 3 % fluorine content, even in the absolute values of $T_N$. Also the run of the curve in the region of superconductivity is similar. The Sm compound shows maximum $T_c$ at a fluorine content of about 11%, which is consistent with values for the La system. Furthermore it becomes clear that all of the published data on the electronic phase diagram are in much better agreement than it previously appeared. When our data is inserted into other published diagrams there is seen to be a close match between our data and data for the Ce compound [7] and the Sm compound [8].

In summary, we can conclude from this investigation of our samples that the substitution of the initial weight with the measured fluorine content leads to a more consistent picture of the electronic phase diagrams.

Acknowledgements. The authors would like to thank M. Deutschmann, J. Werner, R. Müller and S. Müller-Litvanyi for sample preparation work and S. Pichl for WDX analysis.

**Figure Captions**

**Fig. 1**   Fluorine content in $SmO_{1-x}F_xFeAs$ determined by WDX. The measured values deviate from the theoretical line (red color online)

**Fig. 2**   Fluorine content in $LaO_{1-x}F_xFeAs$ determined by WDX. The measured values are in good agreement with the theoretical line (red color online) except for $x < 0.05$ and $x = 0.15$

**Fig. 3**   Electronic phase diagram of $SmO_{1-x}F_xFeAs$ using the nominal (*small symbols*, *thin line*) and measured (*bold symbols* and *lines*, colored online) fluorine content using resistivity data from [5]

**Fig. 4**   Electronic phase diagram of $LaO_{1-x}F_xFeAs$ from [6]. Only the deviant fluorine values at $x < 0.05$ (*bold symbols* and *line*, colored online) are added

**Fig. 1**

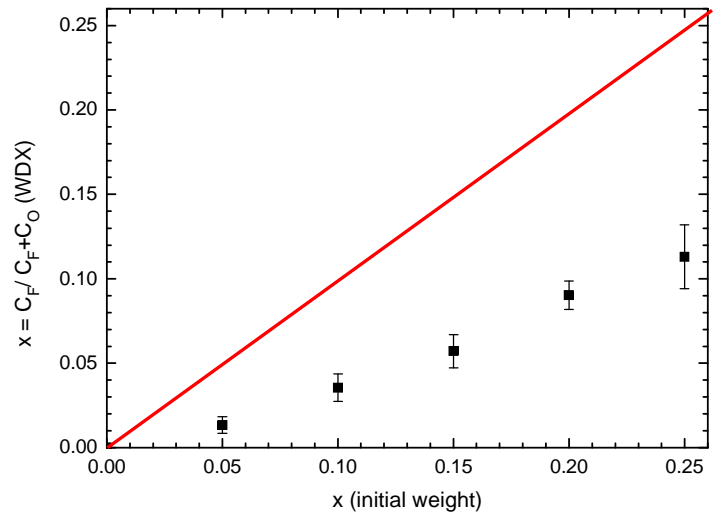

**Fig. 2**

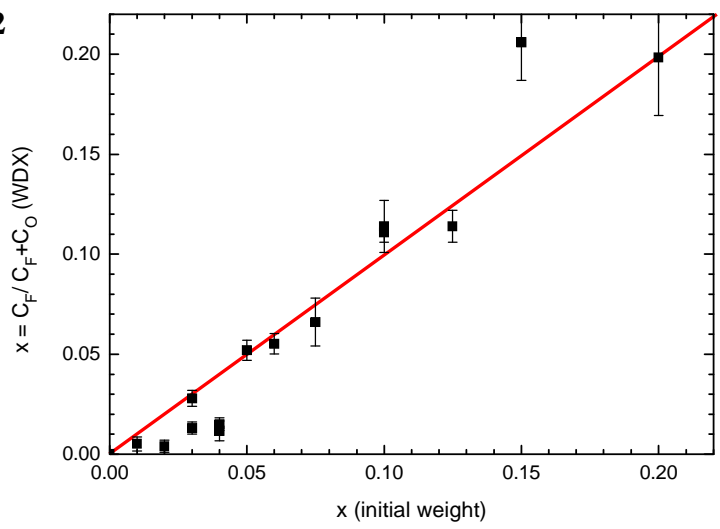

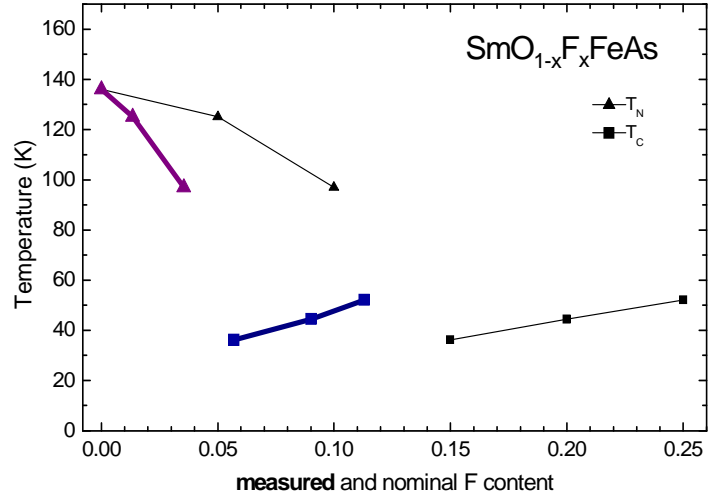

Fig. 3

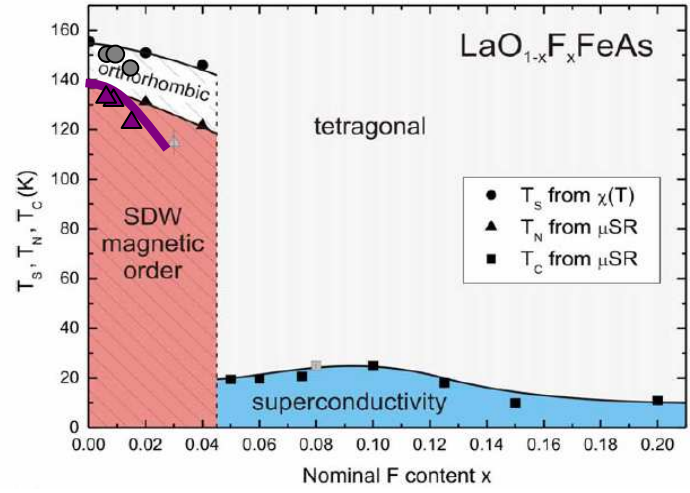

Fig. 4